\documentstyle[12pt,titlepage]{article}
\newcommand{\seqnoll}{\setcounter{equation}{0}}

\newcommand{\Eqref}[1]{Eq.(\ref{#1})}

\newcommand{\bea}{\begin{eqnarray}}
\newcommand{\eea}{\end{eqnarray}}
\newcommand{\be}{\begin{equation}}
\newcommand{\ee}{\end{equation}}
\newcommand{\bc}{\begin{center}}
\newcommand{\ec}{\end{center}}
\newcommand{\ba}{\begin{array}}
\newcommand{\ea}{\end{array}}
\newcommand{\btab}{\begin{tabular}}
\newcommand{\etab}{\end{tabular}}
\newcommand{\bfig}{\begin{figure}}
\newcommand{\efig}{\end{figure}}
\newcommand{\non}{\nonumber}

\newcommand{\minus}{\!-\!}

\newcommand{\BOX}{\hbox {$\sqcap$ \kern -1em $\sqcup$}}

\newcommand{\im}{{\rm Im}}

\def\bigid{\leavevmode\hbox{\small1\kern-3.8pt\normalsize1}}
\def\id{\leavevmode\hbox{\small1\kern-3.3pt\normalsize1}}
\newcommand{\slask}{\!\!\!/}

\newcommand{\Abs}[1]{\left|#1\right|}

\newcommand{\inv}[1]{\frac{1}{#1}}
\newcommand{\del}{\partial}
\newcommand{\ra}{\rangle}
\newcommand{\la}{\langle}

\newcommand{\Tr}{{\rm Tr\,}}

\newcommand{\bfT}{{\bf T }}

\newcommand{\cL}{{\cal L}}
\newcommand{\cO}{{\cal O}}

\newcommand{\apj}[3]{{\it  Ap. J. }{{\bf #1} {(#2)} {#3}}}
\newcommand{\apjl}[3]{{\it  Ap. J. Lett. }{{\bf #1} {(#2)} {#3}}}
\newcommand{\annp}[3]{{\it  Ann. Phys. (N.Y.) }{{\bf #1} {(#2)} {#3}}}

\newcommand{\np}[3]{{\it  Nucl. Phys. }{{\bf #1} {(#2)} {#3}}}
\newcommand{\pr}[3]{{\it Phys. Rev.}{{ \bf #1} {(#2)} {#3}}}

\newcommand{\pl}[3]{{\it  Phys. Lett. }{{\bf #1} {(#2)} {#3}}}
\newcommand{\prep}[3]{{\it Phys. Rep. }{{\bf #1} {(#2)} {#3}}}

\newcommand{\ptp}[3]{{\it  Prog. Theor. Phys. }{{\bf #1} {(#2)} {#3}}}

\newcommand{\zp}[3]{{\it Z. Phys.} {{\bf #1} {(#2)} {#3}}}

\newcommand{\tesla}{{\rm T}}
\newcommand{\lzeroeff}{\cL^0_{eff}}
\newcommand{\lbmeff}{\cL_{eff}^{\beta,\mu}}
\newcommand{\lbmeffzero}{\cL_{0}^{\beta,\mu}}
\newcommand{\lbmeffone}{\cL_{1}^{\beta,\mu}}
\normalsize

\newcommand{\AUTHORS}{\  {\centering
{\large Per Elmfors}\footnote{Email address: elmfors@nordita.dk.} \\
{\sl NORDITA \\
   Blegdamsvej 17 \\
 DK-2100 Copenhagen \O, Denmark \\ }
{\large
David Persson\footnote{Email address: tfedp@fy.chalmers.se.} and
Bo-Sture Skagerstam}\footnote{Email address
address:tfebss@fy.chalmers.se. Research supported by the Swedish
National Research Council under contract no. 8244-103} \\\
\vspace*{1mm}
   {\sl Institute of Theoretical Physics \\
   Chalmers University of Technology and\\
University of G\"oteborg\\
   S-412 96 G\"oteborg, Sweden\\ }}}

\begin{document}
% ==================================================
%
\large
\thispagestyle{empty}
\begin{flushright} NORDITA--93/35    P \\
                   G\3teborg ITP 92--22 \\
                   April 1993  \end{flushright}
\bc
\normalsize
{\LARGE\bf QED Effective Action at Finite Temperature
      and Density\\}
\ec
\vspace*{0.5cm}
\AUTHORS
%
%\vspace*{0.5cm}
\bc
{\bf Abstract} \\
\ec
{\normalsize
The QED effective action at finite temperature and density is
calculated to all orders in an external homogeneous and
time-independent magnetic
field in the weak coupling limit. The free energy, obtained
explicitly, exhibit the expected de\ Haas -- van\ Alphen oscillations.
An effective coupling at finite temperature and density is derived in
a closed form and is compared with renormalization group results.
}
\newpage
% -----------------------------------------------------------------
\normalsize \setcounter{page}{1}
\bc \section{Introduction}
\seqnoll
\label{intro}
\ec
Large magnetic fields are relevant in a number of physical
systems like supernovas\cite{Ginzburg91}, where
$B={\cal O} (10^{10})\tesla$, neutron stars \cite{ShapiroT83}, where
$B={\cal O} (10^{8})\tesla$, or white
magnetic dwarfs \cite{Angel78} in which case
$B={\cal O}
(10^{4})\tesla$. (As a reference the electron mass in units of
tesla is $m^2=\cO(10^9)\tesla$.) The radiative corrections to the magnetic
moment of a Dirac fermion
has been estimated in the presence of such large magnetic fields and it was
argued that they are extremely small \cite{Skagerstam91,Studenikin90}.
It has recently been shown that a plasma at thermal equilibrium can sustain
large fluctuations of the electromagnetic fields. For instance, in
the primordial Big-Bang
plasma, the
amplitude of magnetic field (zero frequency) fluctuations at the time of
the
primordial nucleosynthesis can be as large as
$B = {\cal O}(10^{10}) \tesla$
\cite{TajimaCSK92}. Other systems with large magnetic fields present are
mergers of massive black holes \cite{NatayanPP92},  where
$B={\cal O} (10^{13})\tesla$ or superconducting strings \cite{Witten85},  where
$B={\cal O} (10^{14})\tesla$ or even larger. At the electroweak
phase transition in
the
very early universe it has, furthermore, been argued that very large magnetic
fields,
$B={\cal O} (10^{19})\tesla$, can be generated due to
gradients in the Higgs field
\cite{Vachaspati91}.

In many of these systems one has to consider the effects of thermal
environments.
Calculation of the QED effective potential, i.e. the free energy, has been
attempted  before
either at finite temperature \cite{Dittrich79,Rojas92} or at
finite chemical potential \cite{ChodosEO90}.  In
the latter case the effective action is unfortunately not complete but the
correct form is
presented here. At finite chemical potential and for sufficiently small
temperatures,
the QED effective action should exhibit a certain periodic dependence of
the the
external field, i.e. the well-known de\ Haas -- van\ Alphen
oscillations in condensed
matter
physics. This was not obtained in Ref.\cite{ChodosEO90}.

Let $\cL_{eff}$ denote the QED effective action for a constant B-field
at finite temperature $T=1/\beta$ and chemical potential $\mu$. We
calculate
this effective action
to all orders in $eB$ but with no virtual photons present, i.e. we
consider the weak coupling limit.
A more detailed analysis will be
presented elsewhere \cite{ElmforsPS93}.

%
%====================================================
\bc\section{Derivation of the Effective Action $\cL_{eff}$}
\seqnoll
\label{derivLB}
\ec
The basic relation we need in order to derive the
one-loop correction to the effective Lagrangian
is the identity
\be
\frac{\del\cL_{eff}}{\del m}= i \Tr S_F(x;x)~~~,
\label{BI}
\ee
where $S_F(x;x')$ is the fermion propagator in the external magnetic
field
and the trace is over spinor indices. It can be constructed from
the solutions of the Dirac equation
$(i\del\slask-eA\slask-m)\psi(x)=0$ in such a way that
\be
\ba{c}
S_F(x;x')=\la 0|\bfT [\psi(x)\overline{\psi}(x')]|0\ra\ .
\label{SF}
\ea
\ee
Equation (\ref{BI}) determines the vacuum part,
$\cL_{1}=\cL_{eff}(T=\mu =0,B)$, of the effective action
which should be added to the tree-level, $\cL_{0}=-B^2/2$. At finite
temperature and
density we simply replace the time-ordered vacuum expectation values
in \Eqref{SF} by a
thermal average.
It can be shown that this replacement corresponds to the conventional
calculational rules of
thermo field dynamics. The solutions of the Dirac equation with an external
constant
magnetic field parallel to the $z-$axis are the standard relativistic
Landau levels with energy spectrum given by
\be
E_n(k_z) = \sqrt{m^2+k^2_z+2eBn}\ ,
\ee
where $n=0,1,2,...$ and $k_z$ is the momentum parallel to the magnetic
field. The construction of the propagator is similar to the zero
temperature case \cite{KobayashiS83} except that the propagating
particles can now be exchanged with the heatbath. We find
\be
\label{SF11}
\Tr S_F(x;x)=  \sum_{n=0}^\infty \int \frac{d\omega dk_y dk_z}{(2\pi)^3}
[\Delta+(\Delta^*-\Delta)f_F(\omega)]\ 2m(I_n+I_{n-1})\ ,
\ee
where we have introduced the scalar propagator
\be
\Delta = \inv{\omega^2-k_z^2-m^2-2eBn+i\epsilon}\ .
\ee
The thermal distribution $f_F(\omega)$ is given by
\be
f_F(\omega) = \frac{\theta(\omega)}{e^{\beta(\omega-\mu)}+1}
+\frac{\theta(-\omega)}{e^{\beta(\mu-\omega)}+1}\ ,
\ee
and we use the notation
\be
I_n = \left( \frac{eB}{\pi} \right)^{1/2} \exp \left[
  -  eB \left( x \minus \frac{k_{y}}{eB} \right)^{2} \right]
 \frac{1}{n!} H_{n}^2 \left[ \sqrt{2eB} \left( x \minus \frac{k_{y}}
 {eB} \right) \right]\ ,
\ee
where the functions $H_n$ are Hermite polynomials, and we define
$I_{-1}=0$. From the propagator in \Eqref{SF11} we get both
the vacuum correction $\cL_{1}$ and a thermal correction
$\lbmeff$. It is well-known that a real-time formalism at
finite temperature requires a doubling of the degrees of
freedom and it can be shown that \Eqref{SF11} is the 11-component
of the matrix propagator in thermo field dynamics \cite{UmezawaMT82}.
Here we only need the 11-component for the one-loop calculation.
The vacuum part  of \Eqref{SF11} that survives when
$f_F(\omega)\rightarrow 0$ reproduces the old result by Schwinger
\cite{Schwinger51}
%
%, i.e. ${\cal L}_{1} = {\cal L}_{0} +{\cal L}_{1} $ ,
%where ${\cal L}_{0} = -B^{2}/2$ and
%
\be
{\cal L}_{1} = -\frac{1}{8\pi^{2}} \int_{0}^{\infty}
 \frac{ds}{s^3}exp(-m^{2}s) \left(esB\coth(esB) -1 -
\frac{1}{3}(esB)^2\right)~~.
\ee
Here ${\cal L}_{1}$ has been renormalized  by adding
a second order polynomial in $eB$. We stress that the physics behind
this renormalization is related to the fact that the coefficient in
front of the quadratic term is proportional to the square of inverse
(bare) coupling. This renormalization corresponds to a charge
renormalization as well as a wave function renormalization in such a way
that $eB$ is invariant. This charge renormalization also leads to the weak
coupling
expansion of the QED $\beta$-function, i.e.
\be
\lambda\frac{d}{d\lambda}\alpha(\lambda) = \beta (\alpha(\lambda)) =
\frac{2}{3\pi}\alpha^{2}(\lambda) + {\cal O}(\alpha^{3}(\lambda))~~~,
\label{BETAF}
\ee
where $\lambda$ is a momentum scale factor.
In order to calculate the thermal part $\lbmeff$ of the effective action, we
have to be
careful with the convergence and the analytical structure. We therefore
let the sum over the quantum number $n$ only go to a finite $N$ and take
the limit $N\rightarrow\infty$ at the end. This gives
\be
\label{SFbmu}
\Tr S_F^{\beta,\mu}= \lim_{N\rightarrow\infty} i\,\frac{mB}{\pi^{3/2}}
\,\im\int_{-\infty}^{\infty}\frac{d\omega}{2\pi}
\int_0^\infty\frac{ds}{s^{1/2}}e^{i\frac{3\pi}{4}}
e^{-is(\omega^2-m^2-i\epsilon)} \left[\frac{1+e^{i2sB}}{1-e^{i2sB}}-
\frac{2 e^{i2NsB}}{1-e^{i2sB}}\right]\ .
\ee
The poles in the last factor cancel for finite
$N$, and we cannot let $N\rightarrow\infty$ in a naive
way before deforming the $s$ integration contour to the
imaginary axis. After integrating \Eqref{SFbmu} with
respect to $m$, to get $\lbmeff$, and being careful
with the convergence when deforming the  contours of integration we
arrive at $\lbmeff =\lbmeffzero +\lbmeffone$, where
\be
\lbmeffzero =
\inv{3\pi^2}\int_{-\infty}^\infty d\omega
\theta(\omega^2-m^2)f_F(\omega)(\omega^2-m^2)^{3/2} \ ,
\ee
is the ideal gas contribution in absence of the external field $B$,
and
\bea
\label{Lbmueff}
\lbmeffone &=&
\int_{-\infty}^\infty d\omega
\theta(\omega^2-m^2)f_F(\omega)
\Biggl[\inv{4\pi^{5/2}}\int_0^\infty\frac{ds}{s^{5/2}}e^{-s(\omega^2-m^2)}
(seB\coth (seB) -1)\non\\
 &&-
\inv{2\pi^3}\sum_{n=1}^\infty \left(\frac{eB}{n}\right)^{3/2}
\sin\left(\frac{\pi}{4}-\frac{\pi n}{eB}(\omega^2-m^2)\right) \Biggr]\ .
\eea
This is the main result of our paper. The term with the
sum over $n$ was neglected in Ref.\cite{ChodosEO90} and we
show in Section~\ref{physical}  that it is essential to keep
this term in order to get the
correct physical result.

The finite temperature part of the effective action is directly related to the
free energy of a gas of relativistic fermions in a constant $B$-field.
If $Z(B,T,\mu)$ is the corresponding partition function, without the
contribution from the thermal photon gas, we can also write
\bea
\label{F}
\lbmeff=\frac{\log Z(B,T,\mu)}{\beta V}
&=&\frac{eB}{\beta(2\pi)^2}\sum_{\lambda=1}^2\sum_{n=0}^\infty
\int_{-\infty}^\infty dk\left\{\log(1+e^{-\beta(E_{\lambda,n}-\mu)})\right.
\nonumber \\
&+&
\left. \log(1+e^{-\beta(E_{\lambda,n}+\mu)})\right\}\, ,
\eea
where $E_{\lambda,n}=\sqrt{m^2+k^2+2eB(n+\lambda-1)}$,
and $\lambda$ labels the spin of the fermions.
For $\Abs{\mu} \leq m$,  \Eqref{F} can be rewritten in the
physically less transparent way
\be
\label{Ditt}
\lbmeff=\frac{1}{(2\pi)^2}
\sum_{l=1}^\infty (-1)^l\int_0^\infty
\frac{ds}{s^3}\exp(-\frac{\beta^2l^2}{4s}-m^2s)
eBs\coth(seB)\frac{\cosh(\beta l\mu)}{2}\ ,
\ee
which for $\mu = 0 $ also is an equation given in \cite{Dittrich79}.
 However, it
is not obvious, when written in this form, to see how
 to extract the physical
contents, and how to generalize $\lbmeff$
to $\Abs{\mu}\geq m$, since then it appears to be divergent.
In particular we notice that the high $T$ behaviour given in
\cite{Dittrich79} is not correct. After a Poisson resummation in $l$,
rewriting the sum over $l$ as a contour integral and and carefully
deforming the contours it is, however, possible to show that
\Eqref{Ditt} is equal to \Eqref{Lbmueff} which, of course, is valid for all
$T$ and $\mu$.
%
%===============================================================
\bc
\section{The Physical Content of $\cL_{eff}$}
\seqnoll
\label{physical}
\ec
There are several dimensionful parameters related to $\cL_{eff}$, i.e.
$T,\ \mu,\ m$, and $B$, that can be large or small
compared to each other. We shall only focus on a few
of these limits which we think are particularly
interesting.

The second term in \Eqref{Lbmueff} has an oscillatory
behaviour that we can explore in the limit where
$\{T=0,eB\ll \mu^2-m^2\ll m^2\}$. This is a non-relativistic
limit (in the sense that the kinetic energy is much smaller
than $m$) with a degenerate Fermi sea and a weak external field.
The oscillating part $\cL_{osc}$ of $\lbmeffone$ can be integrated in this
approximation for which we obtain
\be
\label{Losc}
\cL_{osc}=-\frac{(eB)^{5/2}}{4\pi^4 m}\sum_{n=1}^\infty
\inv{n^{5/2}}\cos\left(\frac{\pi}{4}-n\pi\frac{\mu^2-m^2}{eB}\right)\ .
\ee
The oscillation frequency of this periodic function
agrees with the one derived by Onsager \cite{Onsager52}
for the de\ Haas -- van\ Alphen effect. Equation (\ref{Lbmueff}) describes
the full relativistic generalization of this effect. The distance
between the magnetic field of two
adjacent minima of the magnetization is determined
by
\be
\Abs{\inv{eB_i}-\inv{eB_{i+1}}}= \frac{2\pi}{A}\ ,
\ee
where $A$ is the area of an extremal cross section of the
Fermi sea.

In the limit of strong field, $\{eB\gg T^2,m^2,\mu^2-m^2\}$,
we can see from \Eqref{F} that only the lowest Landau level
contribute and $\lbmeff$ goes like a linear function of $eB$.
We shall now reproduce this result from \Eqref{Lbmueff}
and it turns out to be rather non--trivial. The leading $B$
dependence in the first term in \Eqref{Lbmueff} is obtained by
scaling out $eB$ and taking $eB\rightarrow \infty $ in the remainder.
The total contribution is, apart from the thermal integration,

\be
\frac{(eB)^{3/2}}{4\pi^{5/2}}\left[
\int_0^\infty \frac{dx}{x^{5/2}}(x\coth x -1)-\sqrt{\frac{2}{\pi}}
\sum_{n=1}^\infty\inv{n^{3/2}}\right]\ ,
\ee
but this is actually identically zero. The next subleading
term can be shown to be
\be
\lbmeffone = \frac{eB}{2\pi^2}\int_{-\infty}^\infty
d\omega\theta(\omega^2-m^2)f_F(\omega) \sqrt{\omega^2-m^2}~~~,
\ee
which is exactly the leading term from \Eqref{F}. This
calculation shows that the oscillatory term in \Eqref{Lbmueff}
is absolutely necessary to cancel the $B^{3/2}$ term and to give
the correct linear term.

Having shown that the thermal corrections in \Eqref{Lbmueff}
are correct and comprehensible in physical terms, we now
address the question of when they are important, i.e. when
they dominate over the vacuum correction. For
$eB\gg m^2,T^2,\mu^2$, the vacuum correction goes like
\be
\label{LARGEB}
{\cal L}_{1} \approx \frac{(eB)^2}{24\pi^2}\log\left(\frac{eB}{m^2}\right)\ ,
\ee
and it dominates over $\lbmeffone$. However, when
$\{T=0,eB\ll \mu^2-m^2\ll m^2\}$, we have
\be
\label{SMALLB}
{\cal L}_{1} \approx \frac{(eB)^2}{360\pi^2}
\left(\frac{eB}{m^2}\right)^2\ ,
\ee
and
\be
\lbmeffone \approx \frac{(eB)^2}{12\pi^2}\log \left(\frac{\Abs{\mu}}{m}
      +\sqrt{\frac{\mu^2}{m^2}-1}\, \right)
      \approx \frac{(eB)^2}{12\pi^2}
\left(\frac{3\pi^2 n}{m^3}\right)^{1/3}\ ,
\ee
where $en$ is the charge density, and where we have neglected $\cL_{osc}$.
The density correction $\lbmeffone$
therefore dominates over ${\cal L}_{1}$ when
\be
\left(\frac{ n}{m^3}\right)^{1/3} \gg
\inv{30(3\pi^2)^{1/3}}\left(\frac{eB}{m^2}\right)^2\ .
\ee
When $T^2\gg m^2\gg eB$, we have that
\be
\lbmeffone \approx \frac{(eB)^2}{24\pi^2}\log\left( \frac{T^2}{m^2}\right)\ ,
\ee
and we do not agree with the high temperature and weak field
limit in \cite{Dittrich79}.
(We notice the similarity of our result with $\lzeroeff$ for
$eB\gg m^2$.) In this case the thermal contribution  $\lbmeffone$
dominates over ${\cal L}_{1}$ as  given by Eq.(\ref{SMALLB}) when
\be
\frac{T}{m}\gg \exp\left[\inv{30}\left(\frac{eB}{m^2}\right)^2\right]
\approx 1\ .
\ee

Another useful way of extracting the physical information from
$\cL_{eff}$ is to define an effective coupling constant as
\cite{Schwinger51,ChodosOS88}
\be
\label{effalp}
\frac{1}{\alpha(T,\mu,B)}=\inv{\alpha}
-\frac{1}{\alpha B}\frac{\del \cL_{eff}}{\del B}\ ,
\ee
in analogy with the definition of the renormalized
coupling in the vacuum sector
in connection with \Eqref{BETAF}. Special care has
to be taken when evaluating the
derivative
of the oscillating term in Eq.(\ref{Lbmueff}).
In the limit when $eB=0$, we obtain the effective coupling
$\alpha(T,\mu) = \alpha(T,\mu,B=0)$ given by
\be
\label{AlphaTmu}
\frac{1}{\alpha(T,\mu)}= \frac{1}{\alpha } - \frac{2}{3\pi}\int
_{-\infty}^{\infty}d\omega
\frac{\theta(\omega^{2}-m^{2})}{
\sqrt{\omega^{2}-m^{2}}}f_{F}(\omega)~~~.
\ee
When $T=0$, we therefore get an effective coupling
$\alpha(\mu) = \alpha(T=0,\mu)$ such that
\be
\label{Alphamu}
\frac{1}{\alpha(\mu)}= \frac{1}{\alpha }
-\frac{2}{3\pi}\log \left( \frac{|\mu|}{m} + \sqrt{\frac{\mu ^{2}}{m^{2}} -
1}~\right)\ .
\ee
In the limit $\mu=0$, we find the following asymptotic behaviour
of the corresponding effective coupling $\alpha(T) = \alpha(T,\mu = 0)$:
\be
\label{AlphaT}
\frac{1}{\alpha(T)} = \frac{1}{\alpha } -\frac{4}{3\pi}
\int
_{\beta m}^{\infty}
\frac{dx}{
\sqrt{x^{2}-(\beta m)^{2}}}\frac{1}{e^{x} + 1}
\approx \frac{1}{\alpha }
-\frac{2}{3\pi}\log \left(\frac{T}{m}\right)~~~,
\ee
for $T \gg m $.
It is now clear that (only) for $\mu \gg m$ and $T \gg m$ the
effective couplings
$\alpha (\mu)$ and $\alpha (T)$ are solutions to the renormalization group
equation (\ref{BETAF}) when $\lambda $ is identified with $\mu$ and $T$
respectively
(see in this context e.g. Refs.\cite{Morley79,Rojas92}).
We also note that Eq.(\ref{LARGEB}) leads to an effective
coupling $\alpha(B) = \alpha(T=0,\mu = 0,B)$
with an asymptotic behaviour
\be
\frac{1}{\alpha(B)} \approx \frac{1}{\alpha }
-\frac{1}{3\pi}\log \left(\frac{eB}{m^{2}}\right)~~~.
\ee
The effective coupling defined in \Eqref{effalp}  can also be extracted
from the residue of the thermal Debye-screened photon propagator
(see Ref.\cite{Morley79}).

We have only considered a few particular limits
in this paper and there are many more to explore in
different physical situations. All information needed to
do that is contained
in \Eqref{Lbmueff}.

%
%==================================================
\bc\section{Conclusions}
\label{concl}
\seqnoll
\ec
We have established the correct form of the one-loop
QED effective action at finite temperature and density
to all orders in a constant external magnetic field, and the
result differs from earlier attempts. From the form of
$\lbmeff$ presented in \Eqref{Lbmueff} we have checked several
limits that can be understood from a physical point of view.
A great advantage  with our expression for $\lbmeff$ is that
the thermal distribution function $f_F(\omega)$ occurs explicitly.
This means that it is easy to study other thermal situations
by simply replacing $f_F(\omega)$ with some other distribution.

The importance of the thermal correction depends on the value
of $B$, $T$ and $\mu$. In many physically interesting cases
they are all large compared to $m$ and often of the same
order of magnitude, which makes it difficult to obtain
analytical approximations. It is, however, straightforward
to use \Eqref{Lbmueff} for numerical calculations.

Even though the correction to the free energy is small
compared to the value without the external field there
are other quantities that are effected by the presence of
the heatbath. For instance, the magnetization of a degenerate
Fermi sea as was briefly discussed
in Section~\ref{physical}. One could also expect that
QED radiative corrections at
finite
temperature and density and with the strong magnetic fields
discussed in the Introduction could
effect the
electroweak transition rates, relevant for the Big-Bang primordial
nucleosynthesis. We will
return to this issue elsewhere.

We have, furthermore, calculated an effective coupling constant
defined from the part of $\lbmeff$ which is quadratic in $eB$.
It satisfies asymptotically a naive zero temperature
renormalization group
equation where the renormalization scale is replaced by
$T$, $\mu$ or $\sqrt{eB}$.

%
%=======================================================
%{\bf Acknowledgement}
%\\
%
% -------------------- bibliography ---------------------
%\newpage
%\bibliographystyle{unsrt}
%\addcontentsline{toc}{chapter}{References}
%\bibliography{}

%
%
% ---------------------- Figure Captions ----------------
%
%\section*{\centering Figure Captions}
%
\end{document}